\documentclass[aps,twocolumn, amsmath,showpacs,amssymb,prl]{revtex4-1} 
 
\usepackage{graphicx}% Include figure files 
\usepackage{dcolumn}% Align table columns on decimal point 
\usepackage{bm}% bold math 
\begin{document}

\title{Influence of the
shear viscosity of the quark-gluon plasma on elliptic flow
in ultrarelativistic heavy-ion collisions}
 
\author{H.\ Niemi${}^{a}$, G.S.\ Denicol${}^{b}$, P.\ Huovinen${}^{b}$, 
E.\ Moln\'ar${}^{a,c}$,  and D.H.\ Rischke${}^{a,b}$}

\address{$^{a}$Frankfurt Institute for Advanced Studies,
Ruth-Moufang-Str.\ 1, D-60438 Frankfurt am Main, Germany}

\address{$^{b}$Institut f\"ur Theoretische Physik, Johann Wolfgang
Goethe-Universit\"at, Max-von-Laue-Str.\ 1, D-60438 Frankfurt am Main, Germany}

\address{$^{c}$MTA-KFKI, Research Institute for Particle and Nuclear Physics,
H-1525 Budapest, P.O.Box 49, Hungary}

\begin{abstract} 
We investigate the influence of a temperature-dependent shear viscosity over entropy density ratio 
$\eta/s$ on the transverse momentum spectra and elliptic flow of hadrons in ultrarelativistic heavy-ion collisions.
We find that the elliptic flow in $\sqrt{s_{NN}} = 200$ GeV Au+Au collisions at RHIC is dominated 
by the viscosity in the hadronic phase and in the phase transition region, 
but largely insensitive to the viscosity of the quark-gluon plasma (QGP). At the highest LHC energy, 
the elliptic flow becomes sensitive to the QGP viscosity and insensitive to the hadronic viscosity.
\end{abstract}

 
\maketitle 

Ultrarelativistic heavy-ion collisions at the Relativistic
Heavy-Ion Collider (RHIC) and the Large Hadron Collider (LHC) produce
a hot and dense system of strongly interacting matter \cite{experiments}.
The subsequent expansion of the created matter has been shown to exhibit 
a strong degree of collectivity which reveals itself in the 
transverse momentum ($p_T$) spectra of 
finally observed hadrons. In particular, the observed 
large azimuthal anisotropy of the spectra, quantified by the 
so-called elliptic flow coefficient $v_2$, has been interpreted 
as a signal for the formation of a quark-gluon plasma (QGP) with 
very small viscosity in heavy-ion collisions at RHIC~\cite{Gyulassy:2004zy}.

A first indication for the small viscosity of the 
QGP was the agreement between RHIC data and hydrodynamical
calculations in the perfect-fluid limit, i.e., with zero
viscosity~\cite{Huovinen:2006jp}. 
An analysis of the elliptic flow at RHIC in the
framework of relativistic dissipative 
hydrodynamics was performed in Refs.~\cite{allviscous,Shen:2010uy,Song:2010mg}.
These works indeed indicate that the shear viscosity to entropy density ratio,
$\eta/s$, has to be small in order to keep the agreement
between the hydrodynamic simulations and experimental data.

Presently, most hydrodynamical simulations 
assume a constant, i.e., temperature-independent $\eta/s$.
It has been claimed \cite{Song:2010mg} that, in order to describe 
elliptic flow data, this value cannot be larger than 2.5 times the
lower bound $\eta/s=1/4 \pi$ conjectured
in the framework of the AdS/CFT correspondence \cite{AdSCFT}.
A constant $\eta/s$ is, however, in sharp contrast to the behavior 
observed in common liquids and gases, where $\eta/s$ has a strong 
temperature dependence and, typically, a minimum near phase transitions.
A similar behavior of $\eta/s$ is expected for finite-temperature
matter described by quantum chromodynamics (QCD)
near the transition from hadronic matter to the QGP (the QCD 
phase transition) \cite{Csernai:2006zz}. 

A natural question then is whether the
temperature dependence of $\eta/s$ has an effect on the collective
flow of hadrons in heavy-ion collisions. In this work, we investigate
this question in the framework of relativistic hydrodynamics.
We consider a temperature-dependent $\eta/s$ with a minimum near the QCD phase
transition, and compare the results with those obtained for
a constant $\eta/s$ in either the hadronic phase, or the QGP phase,
or both phases. Note that we do not attempt a detailed fit to the data in order
to extract $\eta/s$. Rather, we are interested in the \emph{qualitative\/}
effects of different parametrizations for $\eta/s$ on hadron spectra
and elliptic flow.

Concerning the elliptic flow in Au+Au collisions at RHIC, we find little
difference whether $\eta/s$ is constant in the QGP phase or strongly increasing with
temperature. In contrast, the elliptic flow values are 
highly sensitive to whether we use a constant or temperature-dependent
$\eta/s$ in the hadronic phase, corroborating the findings of
Refs.\ \cite{Hirano:2005xf,Song:2010aq}.
On the other hand, we find that the sensitivity of the elliptic flow to the
values of $\eta/s$ in the high-temperature QGP increases with
increasing collision
energy, while the sensitivity to the hadronic viscosity
decreases. At the highest LHC energy, the above conclusion for RHIC 
energies is reversed: the finally observable elliptic flow is
dominated by the viscosity of the QGP and largely
insensitive to that of the hadronic phase.

Fluid dynamics is determined by the 
conservation of energy, momentum, and charges like 
baryon number. Here, we are interested in the collective
flow at midrapidity in heavy-ion collisions at RHIC and LHC energies.
Consequently, we may neglect baryon number and
assume longitudinal boost invariance~\cite{Bjorken:1982qr}.
We also need the constitutive relations for the dissipative currents.
Here, we only consider the shear stress tensor
$\pi^{\mu\nu}$, the evolution of which we describe in the approach of Israel and Stewart
\cite{IS},
$\langle D\pi^{\mu\nu}\rangle = 
  \frac{1}{\tau_\pi} \left(2 \eta \sigma^{\mu \nu} - \pi^{\mu\nu} \right)
  -\frac{4}{3}\pi^{\mu\nu}\partial_\lambda u^\lambda,$
where $D=u^\mu \partial_\mu$,
$\sigma^{\mu \nu} = \nabla^{<\mu} u^{\nu>}$, and the
angular brackets $<>$ denote the symmetrized and traceless 
projection, orthogonal to the fluid four-velocity $u^\mu$. 
We have also taken the coefficient of the last term in the massless limit.
For details, see Ref.~\cite{Molnar:2009tx}.

We solve the conservation equations numerically by using the 
SHASTA algorithm, see e.g.\ Ref.~\cite{Molnar:2009tx}. The relaxation equations for the components 
of $\pi^{\mu\nu}$ are solved by discretizing spatial gradients 
using centered second-order finite differences. We found that, 
in contrast to SHASTA, this method produces numerically stable 
solutions also for low-density matter at the edges of the system. 

With longitudinal boost invariance, we need to specify the 
values of the energy-momentum tensor in the transverse plane at some 
initial time $\tau_0$. We assume that the initial energy density profile 
is proportional to the density of binary nucleon-nucleon collisions
as calculated from the optical Glauber model (model eBC in Ref.~\cite{Kolb:2001qz}).
The initial transverse velocity and $\pi^{\mu\nu}$ are set to zero. 
The maximum energy densities $\varepsilon_0$ in central collisions (impact parameter $b=0$) are chosen
to reproduce the observed multiplicity in the 0--5\% most
central
$\sqrt{s_{NN}} = 200$ GeV Au+Au collisions at RHIC~\cite{Adler:2003cb} and 
$\sqrt{s_{NN}} = 2.76$ TeV Pb+Pb collisions at LHC~\cite{Aamodt:2010pb}. For the 
$\sqrt{s_{NN}} = 5.5$ TeV Pb+Pb collisions at LHC we use
the multiplicity predicted by the minijet + saturation model~\cite{Eskola:2005ue}. 
The initialization parameters are collected in Table~\ref{tab:ini}.
\begin{table}[t]
\begin{center}
\begin{tabular}{|c|c|c|c|}
\hline
$\sqrt{s_{NN}}$ [GeV] & $\tau_0$ [fm] & $\varepsilon_0$
[GeV/fm$^3$] & $T_{\rm max}$ [MeV] \\ \hline
200	& 1.0	& 24.0	& 335  \\ \hline
2760	& 0.6	& 187.0	& 506  \\ \hline
5500	& 0.6	& 240.0	& 594  \\ \hline
\end{tabular}
\end{center}
\vspace*{-0.4cm}
\caption{\protect\small Initialization parameters for different collisions.}
\vspace{-0.6cm}
\label{tab:ini}
\end{table}

Our equation of state (EoS) is a recent parametrization of lattice-QCD data 
and a hadron resonance gas [s95p-PCE of Ref.~\cite{Huovinen:2009yb}], with 
chemical freeze-out at a temperature 
$T_{\rm chem} = 150$ MeV implemented as in Ref.~\cite{Huovinen:2007xh}.

Hadron spectra are calculated by using the Cooper-Frye freeze-out 
description \cite{Cooper:1974mv} 
with constant decoupling temperature $T_{\rm dec} = 100$ MeV,
which will be shown below 
to give reasonable agreement with both the $p_T$-spectrum
and the elliptic flow coefficient for pions at RHIC.
For the sake of simplicity, we
include viscous corrections to the equilibrium distribution
function $f_0$ as for Boltzmann particles,
even though $f_0$ obeys the appropriate quantum statistics
\cite{Teaney:2003kp}:
\begin{equation}
 f(x,p) = f_0 + \delta f = f_0
\left[1 + \frac{p_\mu p_\nu \pi^{\mu\nu}}{2T^2 (\varepsilon + p)} \right],
\end{equation}
where $p$ is pressure and $p^\mu$ is the hadron four-momentum.
Two- and three-body decays of unstable hadrons are included 
as described in Ref.~\cite{Sollfrank:1991xm}. We include resonances
up to mass $1.7$ GeV.

%%%%%%%%%%%%%%%%%%%%% FIGURE %%%%%%%%%%%%%%%%%%%%%%%%%%%%%%%%
\begin{figure}[bht] 
% \vspace{-0.5cm} 
\includegraphics[width=8cm]{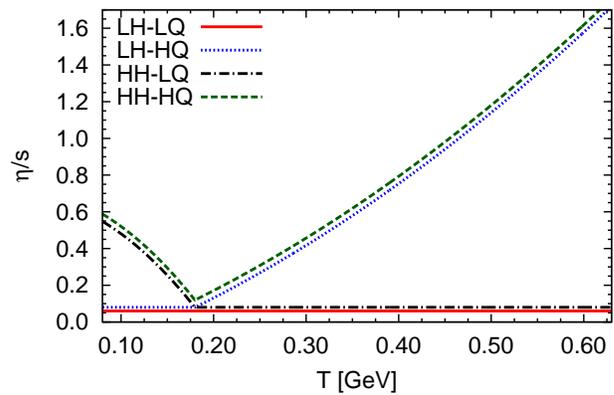} 
\vspace{-0.3cm} 
\caption{\protect\small (Color online) Different parametrizations
of $\eta/s$ as a function of temperature. The \emph{(LH-LQ)} line
is shifted downwards and the \emph{(HH-HQ)} line 
upwards for better visibility.}
\vspace{-0.6cm} 
\label{fig:eta}
\end{figure} 
%%%%%%%%%%%%%%%%%%%%% FIGURE %%%%%%%%%%%%%%%%%%%%%%%%%%%%%%%% 
The shear viscosity to entropy density ratio is
parametrized as follows. For the 
hadronic phase, it reproduces the
results of Ref.~\cite{NoronhaHostler:2008ju}.
In the QGP phase,
$\eta/s$ follows the lattice QCD results of Ref.\
\cite{Nakamura:2004sy}. Then,
$\eta/s$ has to assume a minimum value at a certain temperature;
in our case we take $\eta/s = 0.08$ at $T=180$ MeV.
This is the same parametrization as used in Ref.~\cite{Denicol:2010tr}.
In total we have four cases, see Fig.\ \ref{fig:eta}:
\emph{(LH-LQ)} $\eta/s = 0.08$ for all temperatures,
\emph{(LH-HQ)} $\eta/s = 0.08$ in the hadron gas, and above
$T = 180$ MeV $\eta/s$ increases according to lattice QCD data,
\emph{(HH-LQ)} below $T=180$ MeV, $\eta/s$ is that of a hadron gas,
and above we set $\eta/s = 0.08$,
\emph{(HH-HQ)} we use a realistic parametrization for both
the hadron gas and the QGP.
For the relaxation time we use a result motivated by kinetic
theory $ \tau_\pi = 5 \eta/(\varepsilon+p)$ \cite{Denicol:2010xn}.

\begin{figure*}[ht] 
% \vspace{-0.5cm} 
% \hspace{-0.0cm} 
\includegraphics[width=16.2cm]{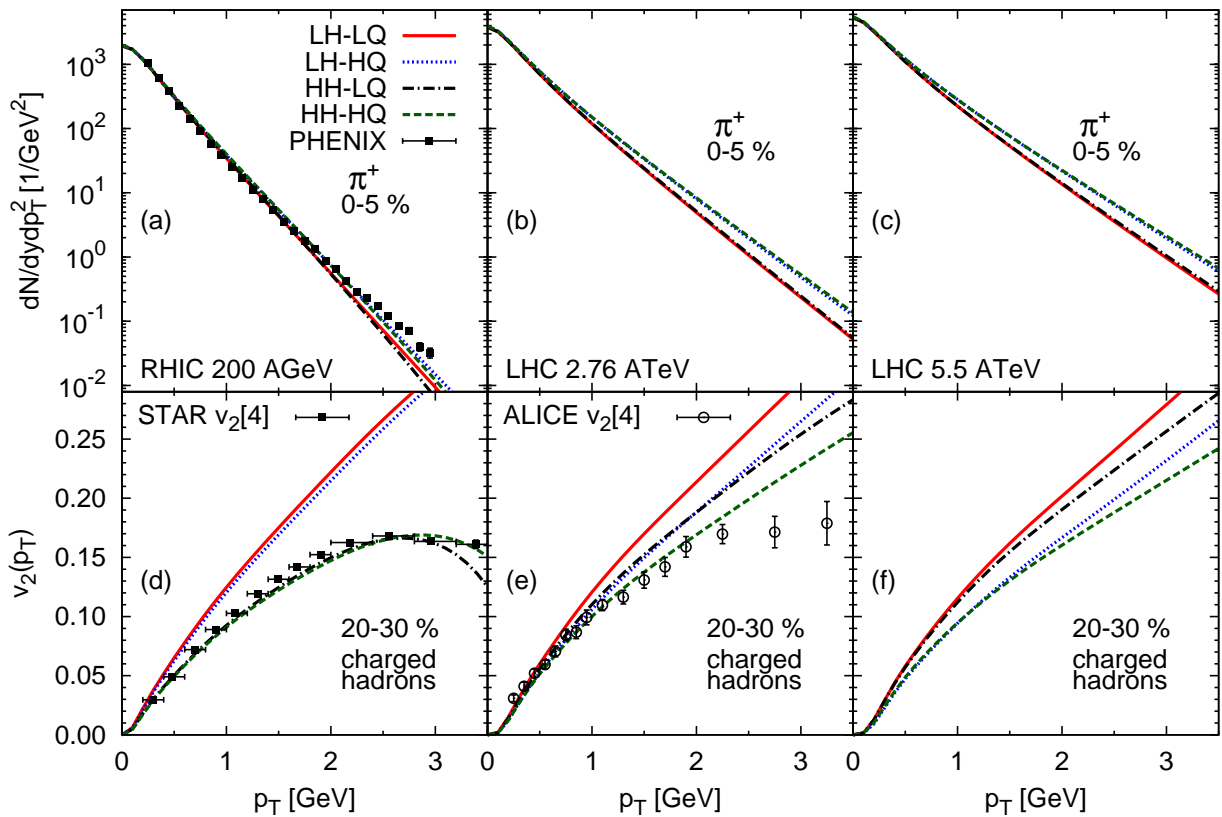} 
 \vspace{-0.3cm} 
\caption{\protect\small (Color online) Transverse momentum spectra of 
positive pions in the 0--5\% most central collisions and elliptic flow
coefficients in the 20--30\% centrality class at RHIC and LHC.
Different curves correspond to the different parametrizations of the 
temperature dependence of $\eta/s$. Data in panel (a) are from 
Ref.~\cite{Adler:2003cb} and in panels (d) and
(e) from Refs.~\cite{Bai,Aamodt:2010pa}.}
\vspace{-0.6cm} 
\label{fig:spectra_all}
\end{figure*}

Figure~\ref{fig:spectra_all}a shows the $p_T$-spectrum of 
positive pions in the 0--5 \% most 
central $\sqrt{s_{NN}}=200$ GeV Au+Au collisions at RHIC. Our 
calculations 
are compared to PHENIX data~\cite{Adler:2003cb}. All the different 
parametrizations of $\eta/s$ give similar agreement with the low-$p_T$ 
pion spectra. For $p_T \agt 1.0$ GeV, 
the parametrizations \emph{(LH-HQ)} and \emph{(HH-HQ)} 
start to give slightly flatter spectra. While the effect of the
QGP viscosity on the $p_T$-slopes is small for our comparatively long initialization time 
$\tau_0=1.0$ fm, it becomes more pronounced for smaller values of
$\tau_0$. On the other hand, the slopes of the spectra are almost independent of the hadronic 
viscosity and this conclusion remains true at least for
$\tau_0=0.2$--$1.0$ fm. 

Figures \ref{fig:spectra_all}b and \ref{fig:spectra_all}c show the 
spectra for $\sqrt{s_{NN}}=2.76$ TeV and $5.5$ TeV Pb+Pb
collisions, respectively. Here we observe a much stronger dependence 
of the $p_T$-spectra on the high-temperature values of $\eta/s$, 
but the main reason for this is the earlier initialization time 
$\tau_0=0.6$ fm. On the other hand, the $p_T$-spectra are 
independent of the hadronic viscosity also at LHC.

%%%%%%%%%%%%%%%%%%%%% FIGURE %%%%%%%%%%%%%%%%%%%%%%%%%%%%%%%%
%%%%%%%%%%%%%%%%%%%%% FIGURE %%%%%%%%%%%%%%%%%%%%%%%%%%%%%%%% 
In Figs.~\ref{fig:spectra_all}d, \ref{fig:spectra_all}e, and 
\ref{fig:spectra_all}f we show the elliptic flow coefficients for 
charged hadrons in the 20--30\% centrality 
class for $\sqrt{s_{NN}}=200$ 
GeV Au+Au collisions and $\sqrt{s_{NN}}=2.76$ TeV and $\sqrt{s_{NN}}=5.5$ TeV
Pb+Pb collisions, respectively. In Fig.~\ref{fig:spectra_all}d the 
results from the hydrodynamic simulations are compared to STAR 4-particle cumulant data 
\cite{Bai} and in Fig.~\ref{fig:spectra_all}e to recent data 
from the ALICE Collaboration~\cite{Aamodt:2010pa}.

We immediately see that, for RHIC, the four parametrizations for
$\eta/s$ produce values for the elliptic flow that fall into
two classes. The curves are largely insensitive to
the values of $\eta/s$ in the QGP phase and follow the value
of the viscosity in the hadron gas:
the parametrizations \emph{(LH-LQ)} and \emph{(LH-HQ)} with constant 
$\eta/s$ in the hadron gas result in larger $v_2(p_T)$ than the 
parametrizations \emph{(HH-LQ)} and \emph{(HH-HQ)} with realistic 
$\eta/s$ in the hadron gas. We have confirmed the insensitivity
to the values of $\eta/s$ in the high-temperature QGP phase by decoupling
the system at $T_{\rm dec}=170$ MeV. In that case, $v_2(p_T)$ is largely 
independent of the $\eta/s$ parametrization. The separation of curves
occurs in the subsequent evolution in the hadronic phase.
This shows that, within this model and at RHIC, viscous effects from
the hadron gas dominate over viscous effects from the QGP, see also
Refs.~\cite{Hirano:2005xf,Song:2010aq}. Due to the strong longitudinal
expansion, the initial shear stress enhances the transverse pressure 
and thus the buildup of the flow anisotropy, but this is counteracted 
by the viscous suppression of anisotropies. Our simulations suggest 
that at RHIC these two effects cancel each other in the QGP phase.

\begin{figure}[bht] 
% \vspace{-0.5cm} 
% \hspace{-0.0cm} 
\includegraphics[width=7.0cm]{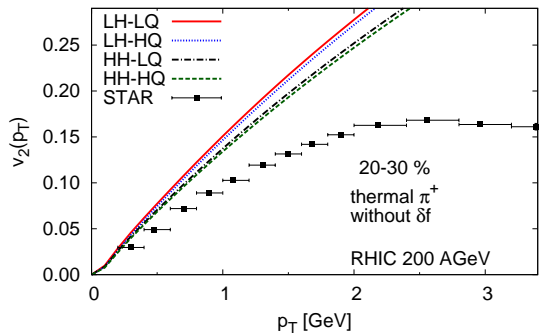} 
\vspace{-0.1cm} 
\caption{\protect\small (Color online) Same as Fig.~\ref{fig:spectra_all}d, 
but for elliptic flow of thermal (i.e., without decays) pions with $\delta f = 0$.}
 \vspace{-0.6cm} 
\label{fig:nodf}
\end{figure}

The main reason for the hadronic suppression of $v_2(p_T)$ are the viscous 
corrections $\delta f$ to the particle distribution function. 
Thus, the values of $\pi^{\mu\nu}$ on the decoupling boundary 
are significantly larger in the case with large hadronic $\eta/s$. 
On the other hand, the azimuthal anisotropies
of the hydrodynamic flow field are quite similar in all cases. 
This is demonstrated in Fig.~\ref{fig:nodf}, where we plot 
$v_2(p_T)$ of pions at RHIC without $\delta f$. All 
curves are much closer to each other, indicating that the space-time evolution
in the hadron gas is similar in all four cases.

We have tested that these conclusions are unchanged if we use different 
$\tau_0 = 0.2$--$1.0$ fm, different EoSs, 
e.g.\ with or without chemical freeze-out, use non-equilibrium initial 
conditions (the same non-zero initial $\pi^{\mu\nu}$ for all four cases), 
or shift the $\eta/s$ parametrizations up by a constant value, such
that $\eta/s$ at $T = 180$ MeV is five times the AdS/CFT lower bound.
Although $v_2(p_T)$ and the slopes of the $p_T$-spectra change when 
we change the setup, the observed sensitivity of $v_2(p_T)$ on the 
viscosity around $T \sim 180$ MeV and below, rather than on the high-temperature 
QGP viscosity is quite generic at RHIC. If we 
increase $\eta/s$ above $T=200$ MeV by a factor of ten in parametrization 
\emph{(HH-LQ)}, the elliptic flow is practically the same as shown 
in Fig.~\ref{fig:spectra_all}d. This confirms
that the value of $\eta/s$ in the high-temperature QGP phase has no effect on the
final observable $v_2(p_T)$ at RHIC, even though during the evolution
the system spends approximately equal times above $T \sim 200$ MeV and 
between $T \sim 170$ and $200$ MeV.

%%%%%%%%%%%%%%%%%%%%% FIGURE %%%%%%%%%%%%%%%%%%%%%%%%%%%%%%%%
%%%%%%%%%%%%%%%%%%%%% FIGURE %%%%%%%%%%%%%%%%%%%%%%%%%%%%%%%% 
Interestingly, the sensitivity of $v_2(p_T)$ to the QGP viscosity increases 
with increasing collision energy, while the sensitivity to the hadronic
viscosity decreases. This can be seen in Figs.~\ref{fig:spectra_all}e
and \ref{fig:spectra_all}f, which show $v_2(p_T)$ for 
$\sqrt{s_{NN}} = 2.76$ TeV and $\sqrt{s_{NN}} = 5.5$ TeV Pb+Pb collisions,
respectively. 

At the highest LHC energy, the behavior of $v_2(p_T)$ is completely
opposite to that at RHIC. It is almost independent of the hadronic viscosity,
but sensitive to the QGP viscosity. In contrast to the RHIC case, at 
LHC the differences in $v_2(p_T)$ are mostly due to the difference in the
transverse flow profiles (caused by the different QGP viscosities)
and not due to the viscous corrections to the distribution function at 
freeze-out. The latter are much smaller than at RHIC: the magnitude of
$\delta f$ is the difference between the curves \emph{(LH-LQ)} and \emph{(HH-LQ)}
or \emph{(LH-HQ)} and \emph{(HH-HQ)} in Fig.~\ref{fig:spectra_all}f.
We have also checked that $v_2(p_T)$ at low-$p_T$ remains insensitive to the hadronic
viscosity, even if we increase the hadronic $\eta/s$ in such way that it reaches
$\eta/s=1.0$ at $T=100$ MeV, but keep the minimum of $\eta/s$ fixed. The collisions 
at $\sqrt{s_{NN}} = 2.76$ TeV are between these two extreme behaviors, as elliptic flow 
depends both on the hadronic and the QGP $\eta/s$.

There are several reasons why the effect of $\eta/s$ on the elliptic flow 
at LHC is so different from that at RHIC: first,
the longer lifetime of the QGP phase, which results in a stronger
dependence of the transverse flow on the viscous properties of the QGP.
Second, once the system decouples, it has much larger transverse
size and velocity gradients are smaller. Subsequently,
dissipative effects from the hadronic stage are smaller and have 
less effect on the observed $v_2(p_T)$.

In conclusion, we have investigated the effects of a temperature-dependent
$\eta/s$ on the hadron spectra and elliptic flow 
coefficients at $\sqrt{s_{NN}}=200$ GeV Au+Au collisions at RHIC
and $\sqrt{s_{NN}}=2.76$ TeV and $\sqrt{s_{NN}}=5.5$ TeV Pb+Pb 
collisions at LHC. We found that in all cases the slopes of the 
$p_T$ spectra of pions depend mainly on the high-temperature $\eta/s$, 
and hardly at all on the hadronic viscosity. 

The effect of $\eta/s$ on the differential elliptic flow $v_2(p_T)$ 
is more subtle. At RHIC energies, $v_2(p_T)$ 
is highly sensitive to the viscosity in hadronic matter
and almost independent of the viscosity in the QGP phase.
In contrast, at the highest LHC energy the opposite holds: elliptic flow 
is almost independent of the hadronic viscosity, but depends strongly 
on the QGP viscosity. Thus the extraction of an $\eta/s$--value for the 
QGP, except for its value at the expected minimum around $T_{c}$, is 
basically impossible using the elliptic flow data at RHIC alone.
On the other hand, a determination of the temperature dependence of
$\eta/s$ in the QGP phase from elliptic flow data seems to be possible at LHC.
This could allow the observation of a possible transition from the strongly coupled plasma 
near $T_c$, see e.g.\ Ref.~\cite{Liao:2006ry}, to the weakly coupled QGP.

This work was supported by the Helmholtz International Center for FAIR within the
framework of the LOEWE program launched by the State of Hesse. The
work of P.H.\ and 
H.N.\ was supported by the Extreme Matter Institute (EMMI).
P.H.\ is also supported by BMBF under contract no.\ 06FY9092.
E.M.\ is supported by OTKA/NKTH 81655 and the AvH foundation.

\end{document}